\documentclass[aps,twocolumn,showpacs,nofootinbib]{revtex4}
\usepackage{graphicx}
\newcommand{\beq}{\begin{equation}}
\newcommand{\eeq}{\end{equation}}
\newcommand{\beqa}{\begin{eqnarray}}
\newcommand{\eeqa}{\end{eqnarray}}

\def\half{\frac{1}{2}}

\def\opone{\leavevmode\hbox{\small1\normalsize\kern-.33em1}}

\begin{document}

\title{On the Impossibility of Covariant Nonlocal "hidden" Variables in Quantum Physics}

\author{Nicolas Gisin \\
\it \small   Group of Applied Physics, University of Geneva, 1211 Geneva 4,    Switzerland}

\date{\small \today}

\begin{abstract}
Local variables can't describe the quantum correlations observed in tests of Bell inequalities. Likewise, we show that nonlocal variables can't describe quantum correlations in a relativistic time-order invariant way.
\end{abstract}

\maketitle

The quantum measurement problem, that is the problem of what makes certain arrangements of atoms and other quantum stuff to act as measurement apparatuses, has led many authors to investigate the possibility to complement quantum physics with additional variables. Until recently this line of research was mostly focussed on local variables and the Bell inequalities that follow as a consequence. But what about the possibility to complement quantum physics with nonlocal variables? Some research concentrated on particular nonlocal variables, as for instance those satisfying a criterion put forward by Prof. Leggett \cite{Leggett}. More generally, it is known since the 1970's that it is not difficult to add nonlocal variables that make quantum physics deterministic \cite{Gudder}. Admittedly, most of these nonlocal models are quite ad-hoc and don't bring much insight into the measurement problem \cite{Bohm}, but the mere possibility of introducing such variables is interesting.

In this note I like to ask whether such additional nonlocal variables could be covariant in the sense of relativistic time-order invariant predictions, that is invariant under a velocity-boost that changes the time ordering of events. We shall see that a pretty simple argument (hence possibly well known to some readers?) reduces any {\it covariant nonlocal variables} to Bell {\it local variables} whose existence is known to be incompatible with well tested quantum predictions. Consequently, since local variables don't exist, likewise covariant nonlocal variables neither exist.

Note that a variable, e.g. $\lambda$, by itself is neither local nor nonlocal; it all depends how the variable $\lambda$ is used in the model. We shall consider nonlocal models in which $\lambda$ determines the measurement outcomes. In the conclusion we'll come back to the question of nonlocal stochastic models.

The argument runs as follows. Consider a situation typical for a test of Bell's inequality, see Fig. 1. Two space-like separated partners, Alice and Bob, each hold a quantum subsystem of an entangled global system, e.g. two 2-level atoms in the singlet state $\psi^{(-)}$. Alice and Bob each have independently the choice of their measurement settings that we label $\vec a$ and $\vec b$, respectively. Let us first look at this situation from a reference frame in which Alice is first to chose her measurement and to secure her result $\alpha$. Assume that her outcome is determined by the usual quantum state and her measurement setting $\vec a$, and by a hypothetical nonlocal variable $\lambda$, presently "hidden" (unknown) to us; her result is thus a function:
\beq\label{alphaAB}
\alpha=F_{AB}(\psi^{(-)}, \vec a, \lambda)
\eeq
where the function's name $F_{AB}$ is chosen to remind us that Alice is First in the time ordering $AB$. Next, in this frame, Bob chooses his measurement setting $\vec b$ and secures his result $\beta$ that is determined by the nonlocal variable $\lambda$:
\beq\label{betaAB}
\beta=S_{AB}(\psi^{(-)}, \vec a,\vec b, \lambda)
\eeq
where $S_{AB}$ stands for Second in the time ordering $AB$. Note that Bob's outcome could depend on Alice's measurement settings $\vec a$; this is the sense in which the variable $\lambda$ together with the function $F_{AB}$ and $S_{AB}$ form a nonlocal model.

\begin{figure}
\includegraphics[width=9cm]{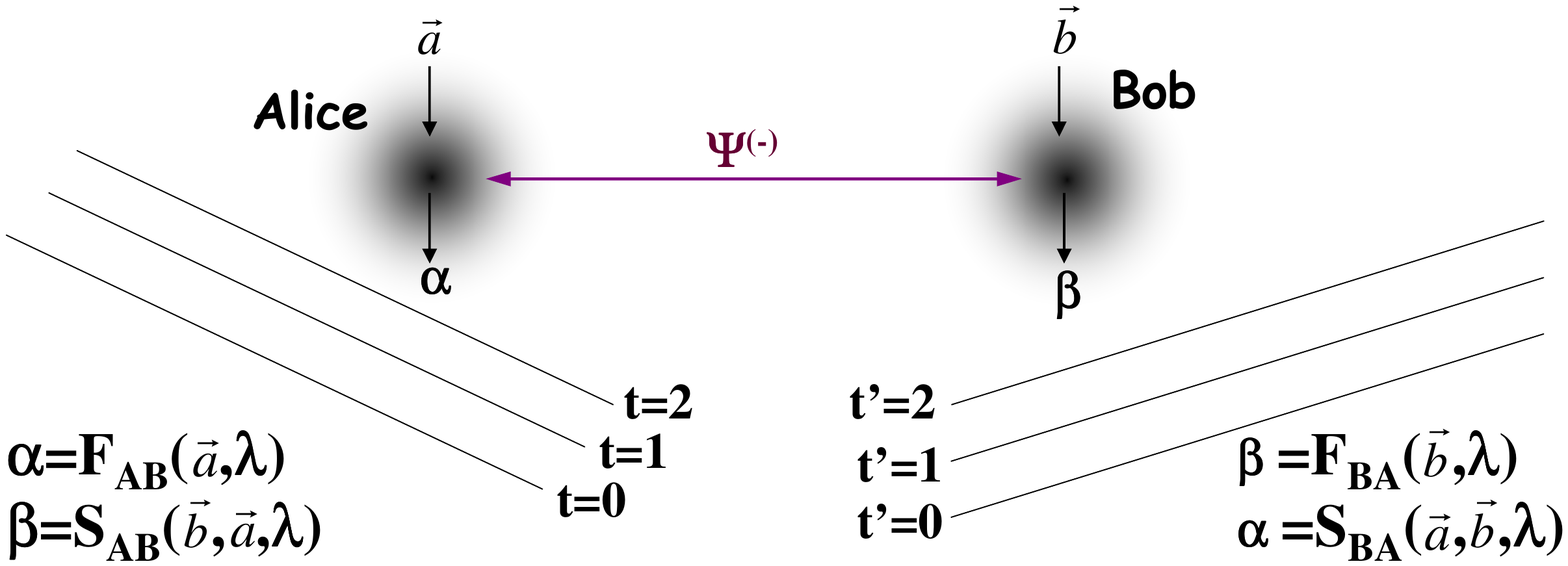}
\caption{\it Standard configuration for Bell inequality tests as seen from two different reference frames. In the first frame, Alice performs her measurement before Bob and her result is determined by the function $F_{AB}$; in the primed frame she is second and her result is given by the function $S_{BA}$, as explained in the text.}
\end{figure}

For example, the following reproduces the standard quantum predictions: $\lambda$ is a pair of real numbers uniformly distributed in the unit square, $\lambda=(r_A,r_B)\in [0..1]\times [0..1]$, $F_{AB}=+1$ iff $r_A\le \half$, and $S_{AB}=+1$ iff $(r_A\le\half$ \& $r_B\le\frac{1-\vec a\vec b}{2})$ or $(r_A>\half$ \& $r_B\le\frac{1+\vec a\vec b}{2})$.

Let us now look at the same experiment from another reference frame, one in which Bob is first, see Fig. 1. The same line of reasoning as above leads to:
\beqa
\beta=F_{BA}(\psi^{(-)}, \vec b, \lambda) \label{betaBA} \\
\alpha=S_{BA}(\psi^{(-)}, \vec a,\vec b, \lambda)  \label{alphaBA}
\eeqa
where a priori the functions $F_{BA}$ and $S_{BA}$ corresponding to the $BA$ time ordering could differ from the functions $F_{AB}$ and $S_{AB}$.

Now, in a covariant nonlocal model Alice's result should be independent of the reference frame, hence from eqs (\ref{alphaAB}) and (\ref{alphaBA}) one gets:
\beq\label{indepy}
\alpha=F_{AB}(\psi^{(-)}, \vec a, \lambda)=S_{BA}(\psi^{(-)}, \vec a,\vec b, \lambda)
\eeq
and similarly for Bob's result $\beta$. From this eq. (\ref{indepy}) one deduces that the function $S_{BA}$ is independent of $\vec b$. But then, the eqs (\ref{betaBA}) and (\ref{alphaBA}) define a local model in the sense of Bell. Hence, any covariant nonlocal model is equivalent to a Bell-local model and, consequently, contradicts well tested quantum predictions, the violation of Bell's inequality.

In conclusion, we have shown that there is no covariant nonlocal models of quantum correlations, not more than local models.

We end this brief note with some comments. First, our argument assumes a deterministic model, i.e. for any given $\lambda$ there is a unique pair of results, one on each side. For stochastic models the situation is interesting. If one interprets probabilities in the usual "classical" way, one may merely add some random variable $\lambda$ to turn the model deterministic, as can be done for any probability appearing in classical physics and as illustrated in the paragraph in between eqs (\ref{betaAB}) and (\ref{betaBA}). But this possibility is ruled out by our argument. Hence, the square of probability amplitudes are not probabilities in the usual sense: they can't be interpreted as mere epistemic probabilities. Consequently, it may help to use a different terminology, like, e.g., quantum propensities. This issue is further discussed in the companion note \cite{FWTrGRWf}.
Next, if the model is also assumed to be nonlocal in time in the sense that the first measurement result depends also on the second input, e.g. $\alpha=F_{AB}(\psi^{(-)}, \vec a, \vec b, \lambda)$, as in Aharonov et al. 2-time model \cite{TwoTime}, then the argument fails (though I should admit that I don't understand what time with two directions could mean).
Thirdly, our little result emphasizes once again the extraordinary robustness of quantum physics against any conceivable change. This may explain why the founding fathers could discover quantum physics based on the very few data available to them at the time: there was simply not much alternative!
Finally, one should mention that a way out of our entire argumentation is to assume the existence of one preferred universal reference frame which determines unequivocally one and only one time ordering for all events.

\small

\section*{Acknowledgment} I benefited from stimulating discussions with Sandu Popescu who convinced me to write down the result presented in this little note.


\begin{thebibliography}{99}
\bibitem{Bohm} With the noticable exception of Bohm's model which led, among other results, to J. Bell discovery of his inequality.
\end{thebibliography}
\end{document}